\documentstyle[12pt,twoside,fleqn,espcrc1,epsfig]{article}

\newcommand{\AmS}{{\protect\the\textfont2
  A\kern-.1667em\lower.5ex\hbox{M}\kern-.125emS}}
\newcommand{ \rts }{$\sqrt{s_{_{\rm NN}}}$ }
\newcommand{ \pt }{$p_t$ }

\title{Hadron freeze-out conditions in high energy nuclear collisions} 
\author{Nu Xu and Masashi Kaneta \\
{\it Lawrence Berkeley National Laboratory, Berkeley, CA 94720, USA}}

\begin{document}
\maketitle

\section{Introduction}

The purpose of the current heavy ion programs at CERN (Switzerland)
and Brookhaven National Laboratory (USA) is to probe strongly
interacting matter under extreme conditions, i.e.  at high densities
and temperatures. The central subject of these studies is the
transition from the quark-gluon plasma to hadronic matter.  In the
early phases of ultra-relativistic heavy ion collisions, when a hot and
dense region is formed in the center of the reaction, there is copious
production of up, down, and strange quarks.  Transverse expansion is
driven by the multiple scattering among the incoming and produced
particles.  As the medium expands and cools, the quarks/gluons combine
to form the hadrons that are eventually observed.
  
In June 2000, the long awaited gold collisions were delivered by the
RHIC collider at \rts = 130 GeV. A great amount of new data, analyzed
from that brief period of running, was presented at this
conference~\cite{vedbeak01,zajc01,roland01,harris01}. From SPS,
results on the systematic of the transverse momentum distributions of
$\phi$~\cite{friese01} and $J/\psi$~\cite{bordalo01,na50plb} were
shown.  New results from 40 and 80 A$\cdot$GeV (\rts $\approx 9$ and
12 GeV) \cite{harry01,blume01} collisions were also shown.

In this paper, we will summarize the recent experimental results on
transverse momentum distributions and particle ratios. The related
physics issues are chemical equilibrium and collective expansion.
While the former has to do with the inelastic collisions, the later is
dominated by the elastic cross section. We will focus at the relatively
low transverse momentum region, \pt $\le 2$ GeV/c, where the bulk
production occurs. For results on high momentum transfer, global
measurements, and correlations readers are referred to
\cite{dress01,steiberg01,panitkin01}.

\section{Kinetic freeze-out: transverse momentum distributions}

At this conference, identified particle transverse momentum
distributions at mid-rapidity were shown by the BRAHMS, PHENIX, and
STAR experiments~\cite{vedbeak01,zajc01,harris01,julia01,mauel01}.
While PHENIX measured $\pi, K,$ and $p$ up to 2.5 GeV/c in \pt, the STAR TPC
provided clean spectra of identified particles up to about 1 GeV/c in
\pt within a somewhat wider rapidity window.  The preliminary
transverse momentum distributions for negative pions, kaons, and
anti-protons from the STAR Collaboration are shown in Fig.\,1. The
results are from the 6\% most central Au+Au collisions. Hydrodynamic
motivated fits are also shown in the figure as dashed
lines~\cite{heinz87}.  The extracted common freeze-out temperature and
collective velocity parameters are about 100 MeV and 0.6$c$,
respectively. In Fig.\,2, the preliminary transverse momentum
distribution of anti-protons from the PHENIX
collaboration~\cite{julia01} is compared to the STAR
results~\cite{mauel01}. It is important to note that the
transverse momentum distributions are consistent within the overlap
region. Below $m_t-mass$ = 0.7 GeV/$c^2$, both sets of data are
consistent with the hydrodynamic fit. The lower part of the
distribution is from the NA44 Collaboration at SPS (\rts = 17.2 GeV)
\cite{na44_coll_exp}.

\hspace{-0.2cm}
{\psfig{figure=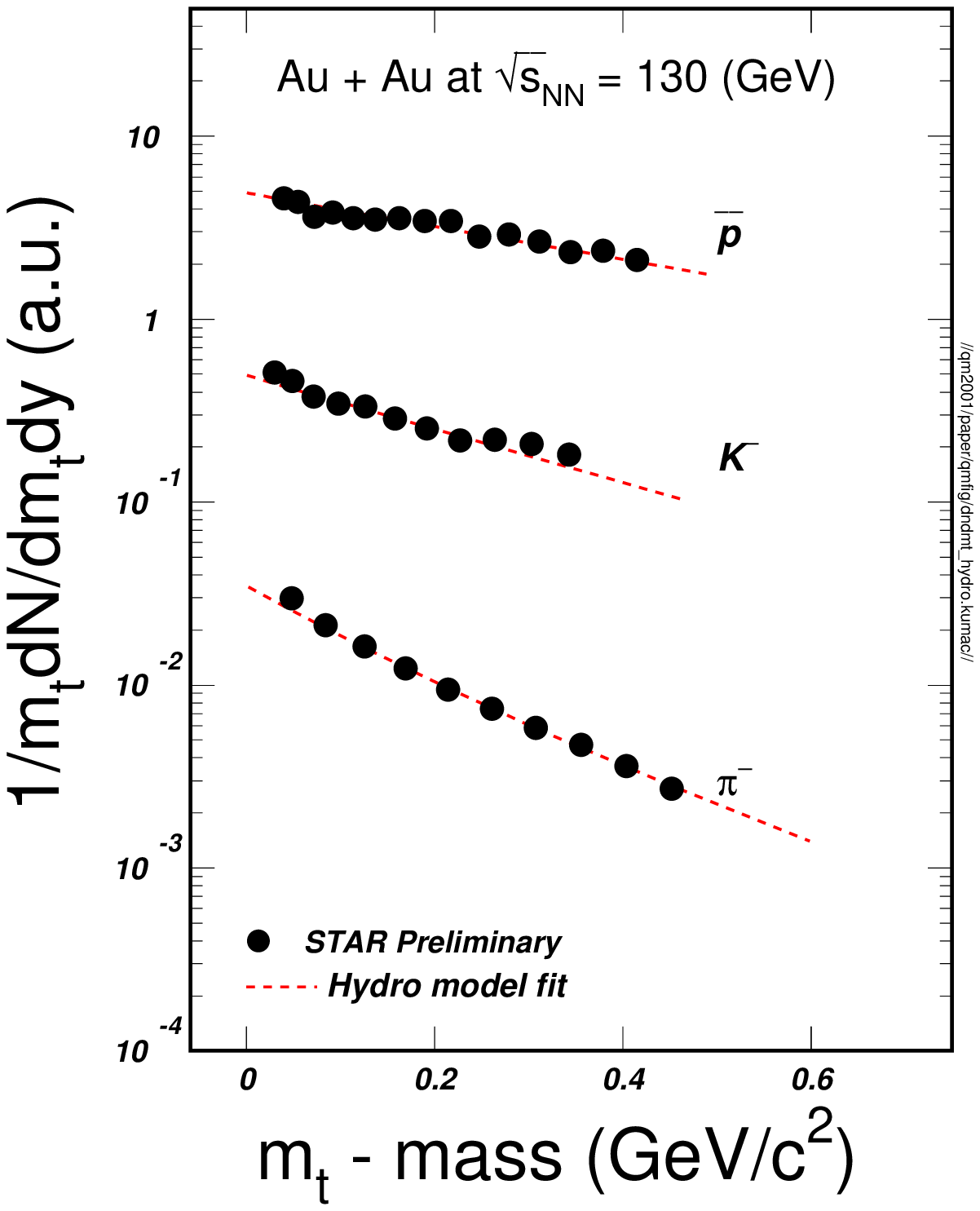,height=8.50cm}}

\vspace{-.20cm} \hspace{-.10cm} 
\begin{minipage}{7cm}
  {\small{{\bf Figure 1 }{\it The STAR preliminary results of the
        transverse mass distributions for negative pions, kaons, and
        anti-protons from Au+Au central collisions. Dashed lines
        represent the model fit results. }}}
\end{minipage} 

\vspace{-10.cm} \hspace{7.65cm}
\begin{minipage}{7.cm}
{\psfig{figure=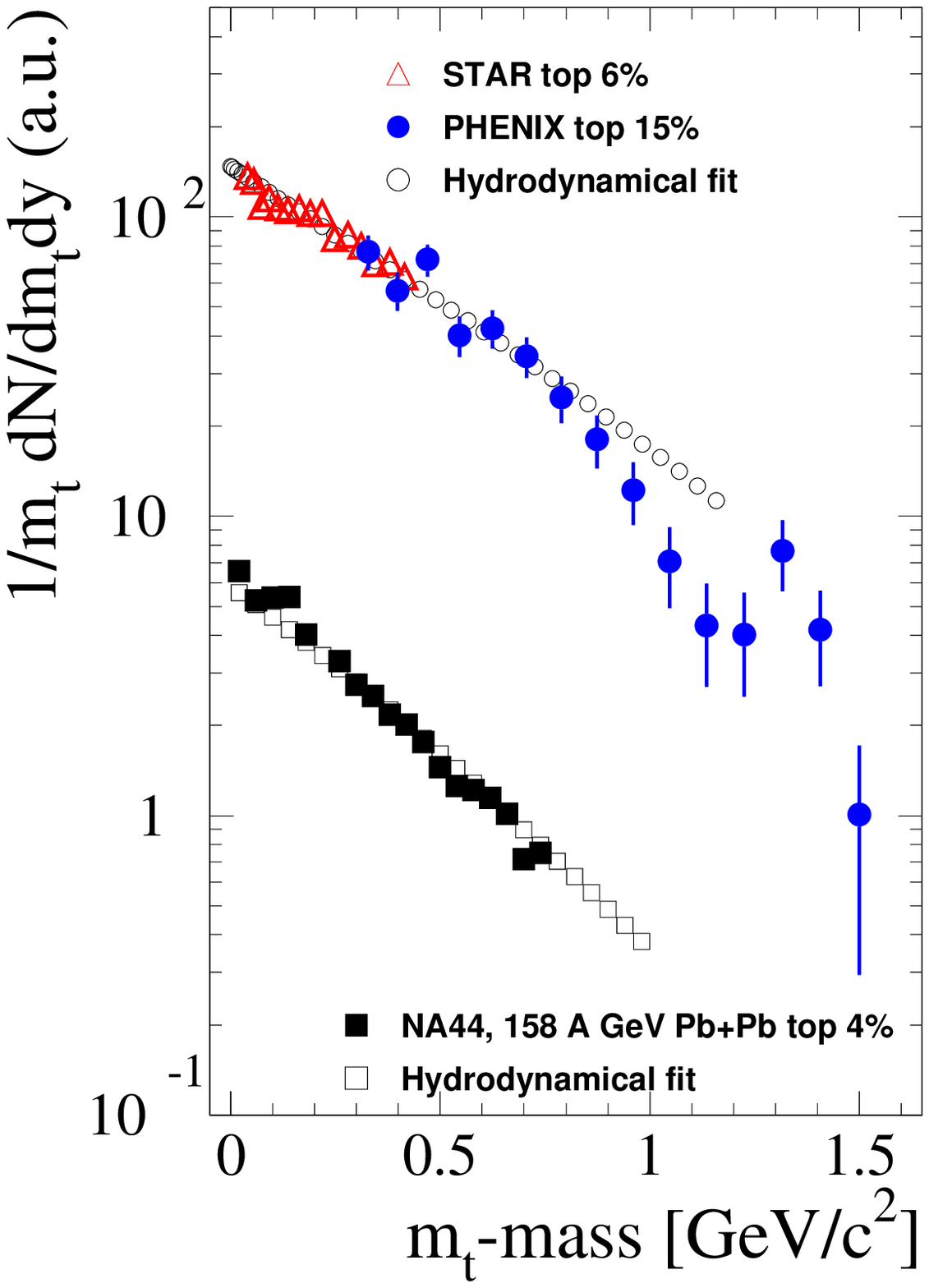,height=7.250cm,width=7.cm}}
\end{minipage}

\vspace{-.35cm} \hspace{8.0cm}  
\begin{minipage}{7.cm}
  {\small{{\bf Figure 2 } {\it Preliminary anti-proton transverse mass
        distributions from STAR and PHENIX (circles) are compared.
        Squares represent the results from central Pb+Pb
        collisions. Open symbols represent the model fit results. }}}
\end{minipage} 

\vspace{0.5cm}

The measured transverse momentum distributions also have been fitted
by the exponential function $f = A \cdot exp(-m_t/T),$ where $T$ is
the slope parameter and $A$ is the normalization constant.  The
magnitude of the slope parameter provides information on temperature
(random motion in local rest frame) and collective transverse flow.
Fig.\,3(a) shows the measured particle slope parameters from Pb+Pb
central collisions at SPS (\rts = 8.8 GeV and \rts = 17.2 GeV)
energies~\cite{harry01,blume01,na44_coll_exp,tsukuba}. The new results are:
\begin{itemize}
\item Particle distributions from 40 A$\cdot$ GeV (\rts = 8.8 GeV)
  collisions were reported by NA45 and NA49~\cite{harry01,blume01}.
  The slope parameters are similar to the ones observed at
  158A$\cdot$ GeV, {\it i.e.}, they follow the established systematic
  trend~\cite{nxuqm96};
  
\item For the first time, the systematic of the transverse momentum
  distributions for charm particles ($J / \psi$) was reported by the
  NA50 experiment~\cite{bordalo01,na50plb}. It is interesting to note
  that the slope parameter of $J / \psi$ is similar to those for
  $\phi$ and $\Omega$;
  
\item For central collisions the $\phi$ slope parameter of NA49 is
  about 300 MeV~\cite{friese01} whereas that of the NA50 is about 240
  MeV.  NA49 and NA50 reconstructed $\phi$ mesons via $K^+K^-$ and
  $\mu^+\mu^-$ channels, respectively.  As discussed in
  \cite{johnson,soffsqm00}, part of the difference may be caused by
  the final state interaction of the decay kaons. In fact, if one
  studies the centrality dependence of the $\phi$ slope parameter, one
  finds that at peripheral collisions, both experimental results agree
  with each other and the value is close to that from $p+p$
  collisions~\cite{friese01,bordalo01}.
\end{itemize}

\vspace{-0.75cm} 
\hspace{2.5cm}
\begin{minipage}{10.5cm}
\centerline{\psfig{figure=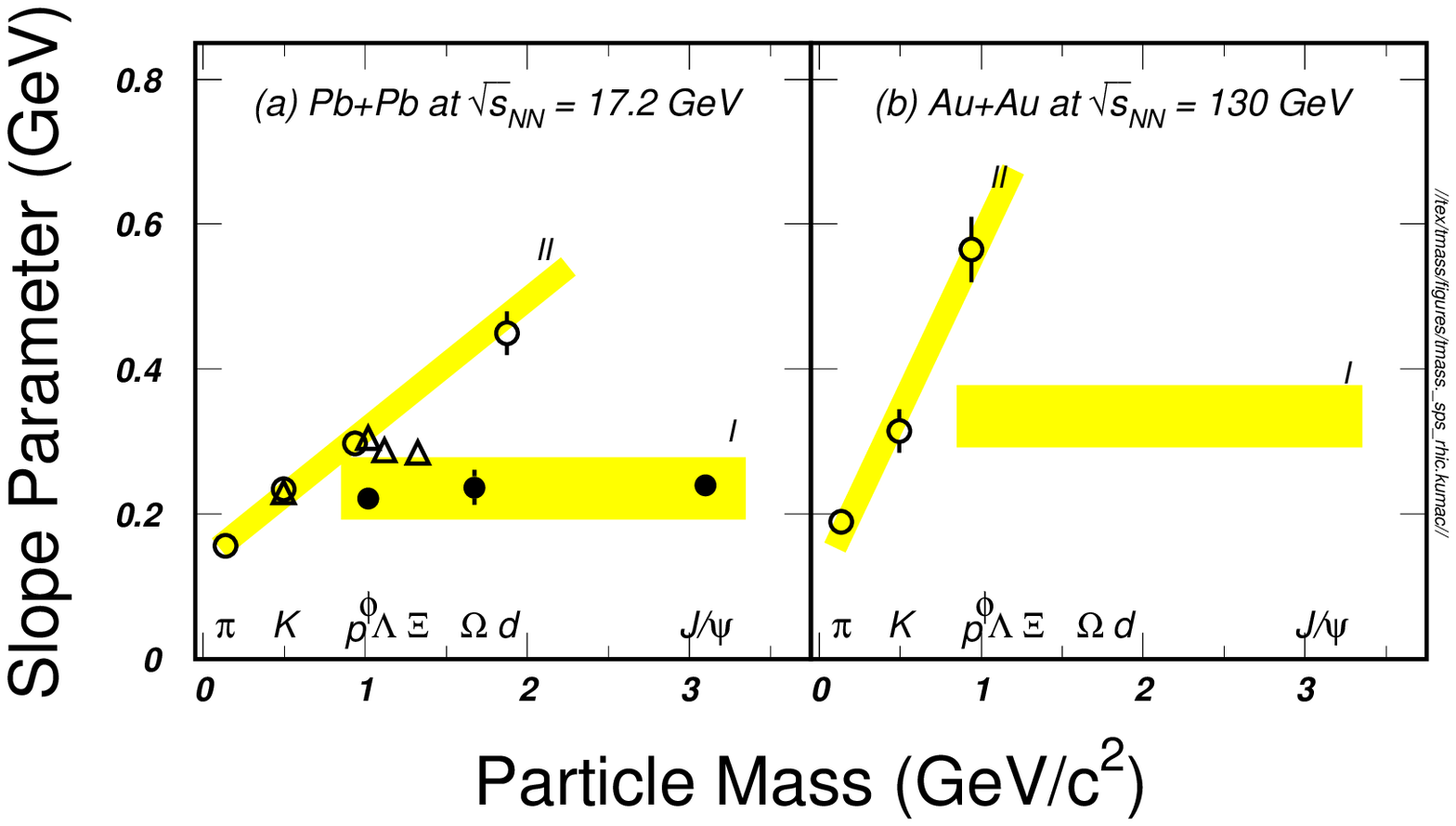,height=10.0cm}}
\end{minipage}

\vspace{-1.cm}
\hspace{.5cm}
\begin{minipage}{14.5cm}
  {\small{{\bf Figure 3 }{\it  Slope parameters as a function of particle
      mass for (a) Pb+Pb central collisions at SPS (\rts  = 17.2 GeV)
      and Au+Au central collisions at (b) RHIC (\rts = 130 GeV).
      Weak and strong interacting limits are indicated by $I$ and
      $II$, respectively.}}}
\end{minipage}

\vspace{.5cm}

Preliminary results of the slope parameters from Au+Au central
collisions at RHIC (\rts = 130 GeV)~\cite{harris01,mauel01} are shown
in Fig.\,3(b).  It is obvious that the mass dependence of the slope
parameter is stronger than that from collisions at SPS energies.

As one can see from Fig.\,3(a), the slope parameters appear to fall
into two groups: group (I) is flat as a function of particle mass,
whereas the slopes in group (II) increase strongly with particle mass.
At the RHIC energy, the slope parameter systematic of $\pi, K,$ and $p$
shows an even stronger dependence on the particle mass.  The strong
energy dependence of the slope parameter might be the result of the
larger pressure gradient at the RHIC energy. With a set of reasonable
initial/freeze-out conditions and equation of state, the stronger
transverse expansion at the RHIC energy was, in deed, predicted by
hydrodynamic calculations \cite{pass01,deark01,passplb01}. In
addition, the results are consistent with the large value of the event
anisotropy parameter $v_2$ at the RHIC
energy~\cite{starflow,snelling01}.

Within the hadronic gas, the interaction cross section for particles
like $\phi, \Omega,$ and $J / \psi$ are smaller than that of $\pi, K,$
and $p$~\cite{hsx98}.  Therefore the interactions between them and the
rest of the system are weak, leading to the flat band behavior in
Fig.\,3(a).  On the other hand, the slope parameter of these weakly
interacting particles may reflect some characteristics of the system
at hadronization. Then it should be sensitive to the strength of the
color field~\cite{schwinger,soff,bleicher001}. Under this assumption,
the fact that the weak interacting particles show a flat slope
parameter as a function of their mass would indicate that the $flow$
develops at a later stage of the collision. Should the collective flow
develop at the partonic level, one would expect a mass dependence of
the slope parameters for all particles \cite{thews01}.

Based on the idea of Color Glass Condensate, McLerran and
Schaffner-Bielich predicted~\cite{mclerran01} that the measured mean
\pt is controlled by the intrinsic \pt broadening rather than
transverse flow. It will be interesting to see the slope parameters of
$\phi, \Omega,$ and $J / \psi$ from collisions at RHIC energies. As
indicated by the horizontal bar in Fig.\,3(b), one might expect both
the absolute magnitude and the slope as a function of particle mass to
be different from that observed at SPS energies.

Figure 4 shows the bombarding energy dependence of the freeze-out
temperature $T_{fo}$ and the average collective velocity $\langle
\beta_t \rangle$. It is interesting to observe that both $T_{fo}$ and
$\langle \beta_t \rangle$ saturate at a beam energy of about 10
A$\cdot$GeV. The saturation temperature is about 100 - 120 MeV, very
close to the mass of the lightest meson~\cite{pomeranchuk}.  The steep
rise of the $T_{fo}$ and $\langle \beta_t \rangle$ up to about 5
A$\cdot$GeV incident energy indicates that at low energy collisions
the thermal energy essentially goes into kinetic degrees of freedom.
The saturation at $\sim$10 A$\cdot$GeV shows that particle generation
becomes important.  As proposed in~\cite{pomeranchuk,heg}, for a pure
hadronic scenario there may be a limiting temperature $T_c \approx
140$ MeV in high-energy collisions, although the underlying physics
for both, the transition from partonic to hadronic degrees of freedom
and the transition from interacting hadrons to free-streaming is not
clear at the moment.  By coupling the limiting temperature idea to a
hydrodynamic model calculation, St\"ocker {\it et al.} successfully
predicted~\cite{stocker} the energy dependence of the freeze-out
temperature. It is worth noting that measurements of the pion phase
density~\cite{cramer01,lisa01,laue01} at freeze-out also
show saturation at $E_{beam} \sim$10 A$\cdot$GeV.

\hspace{.2cm} \vspace{-0.2cm}
{\psfig{figure=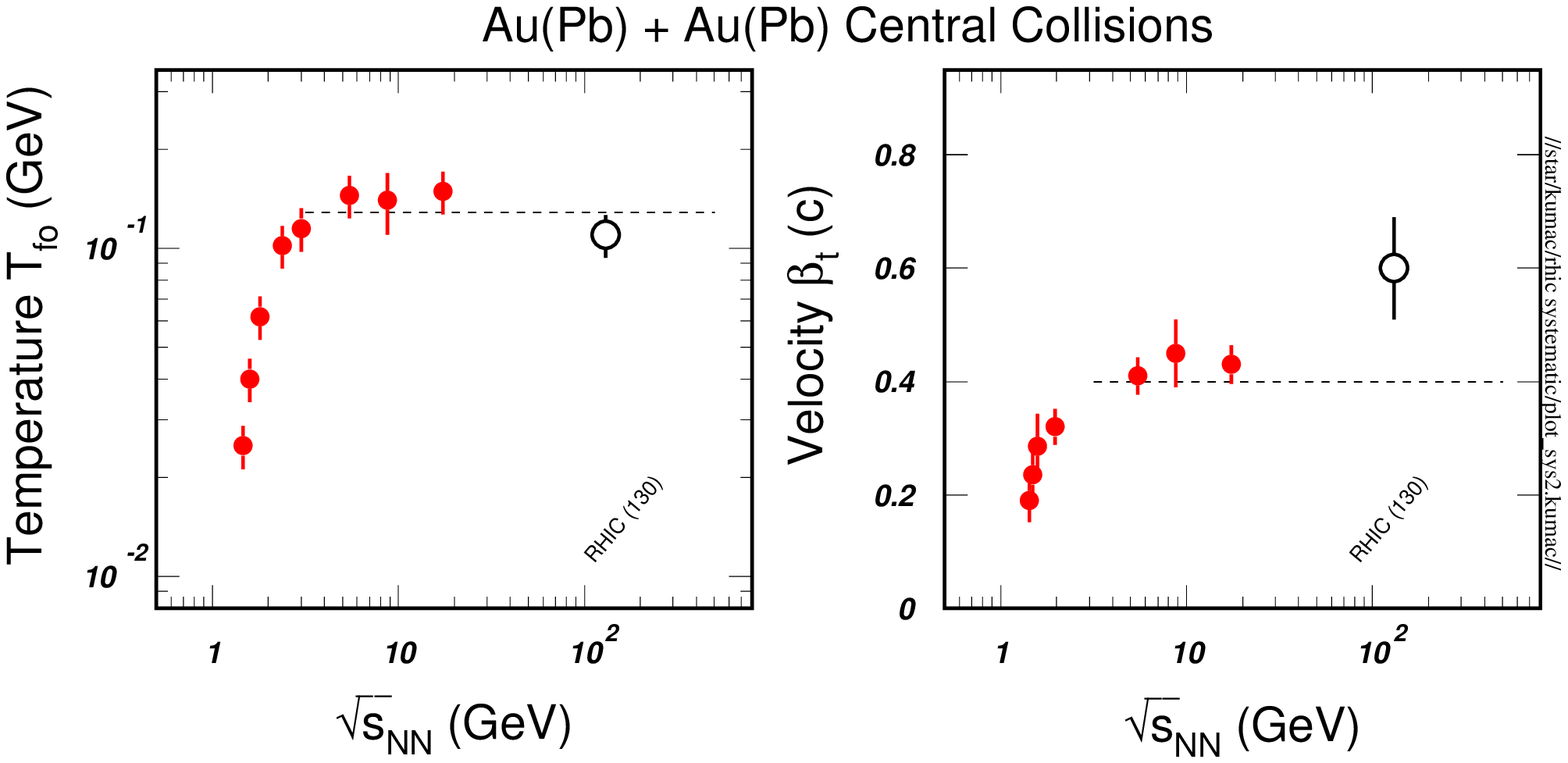,height=7.50cm}}

\vspace{-.50cm} \hspace{-.50cm}  
\begin{minipage}{15.5cm}
  {\small{{\bf Figure 4 }{\it Systematics of kinetic freeze-out
        temperature parameter $T_{fo}$ and average collective
        transverse flow velocity $\langle \beta_t \rangle$ as a
        function of beam energy.  At \rts $\approx$ 5 GeV, both values
        of temperature and velocity parameters seem to saturate.
        However, the velocity parameter extracted from central
        collisions at the RHIC energy is higher than values from
        collisions at lower beam energies.  }}}
\end{minipage} 

\vspace{.5cm}

At the RHIC energy, the collective velocity parameter seems to be
larger than that from collisions at AGS/SPS energies. This can already
be seen in Fig.\,3 where the increase from pion to proton at \rts =
130 GeV (Fig.\,3(b)) is much faster than at \rts = 17 GeV
(Fig.\,3(a)). Is this the manifestation of van Hove's \cite{vH83}
picture? On the other hand, compared to results from lower energy
collisions, the temperature parameters seem to be lower. Is this the
consequence of hydrodynamic expansion~\cite{navarra92} from a higher
initial density fireball? An energy scan between \rts = 20 - 130 GeV
will be extremely important in order to study this evolution in more
detail.

\section{Mid-rapidity $\bar{p}/p$ ratios and net-proton distributions}

All RHIC experiments have consistently measured the ratio of
$\bar{p}/p$ at mid-rapidity
\cite{starpbarp,beadern01,georg01,hirok01,huan01}. The dependence of
the $\bar{p}/p$ ratios as a function of the center of mass energy is
shown in Fig.\,5. Only central collision results were used for heavy
ion collisions. Clearly, although the ratio is not unity, a dramatic
increase in the ratio is observed from SPS to RHIC, meaning that
the system created at \rts = 130 GeV is close to net-baryon free.

\vspace{-0.5cm}
\hspace{-4.25cm}
\centerline{\psfig{figure=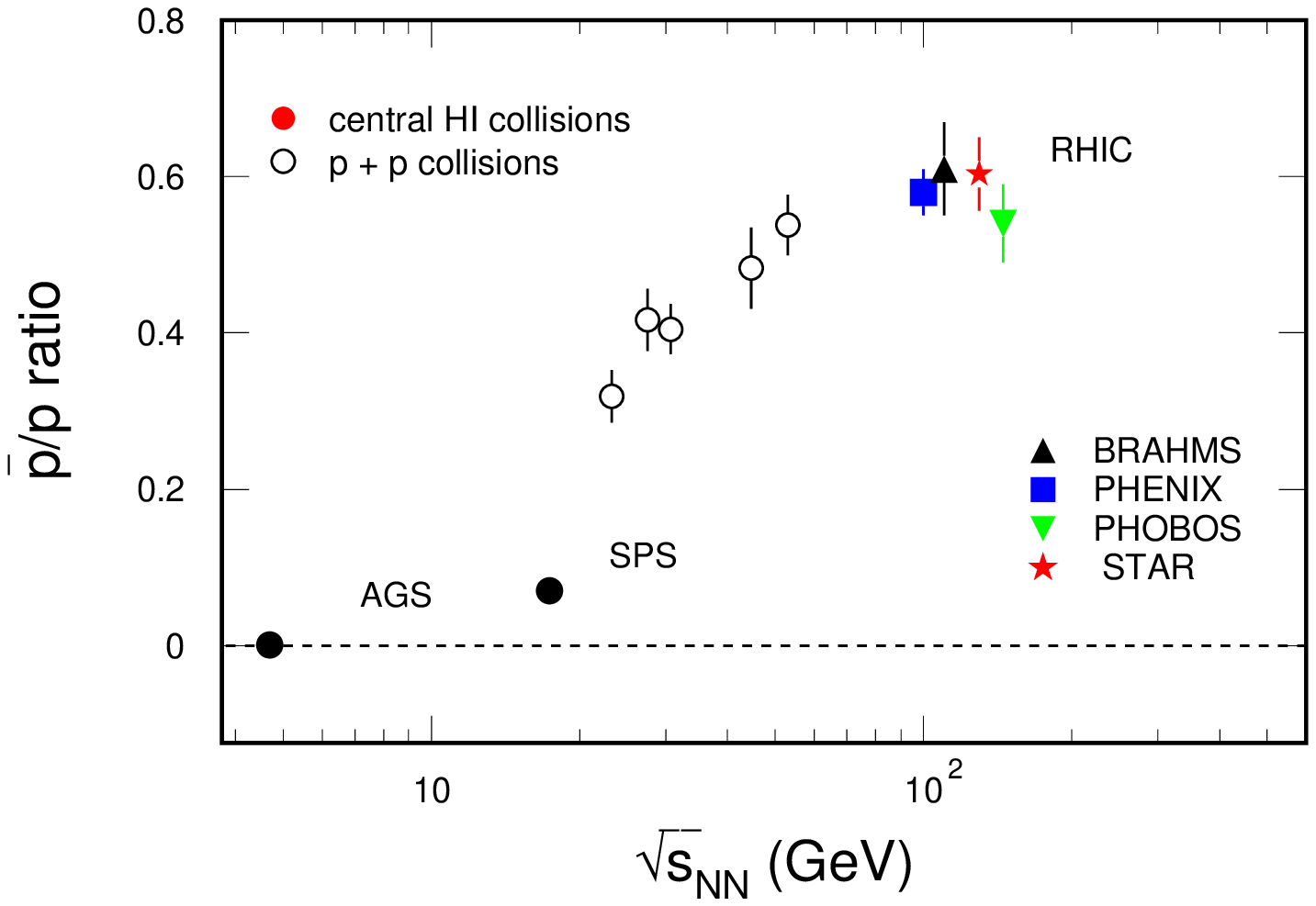,height=8.0cm}}

\vspace{-8.cm} \hspace{8.25cm}
\begin{minipage}{8.0cm}
\centerline{\psfig{figure=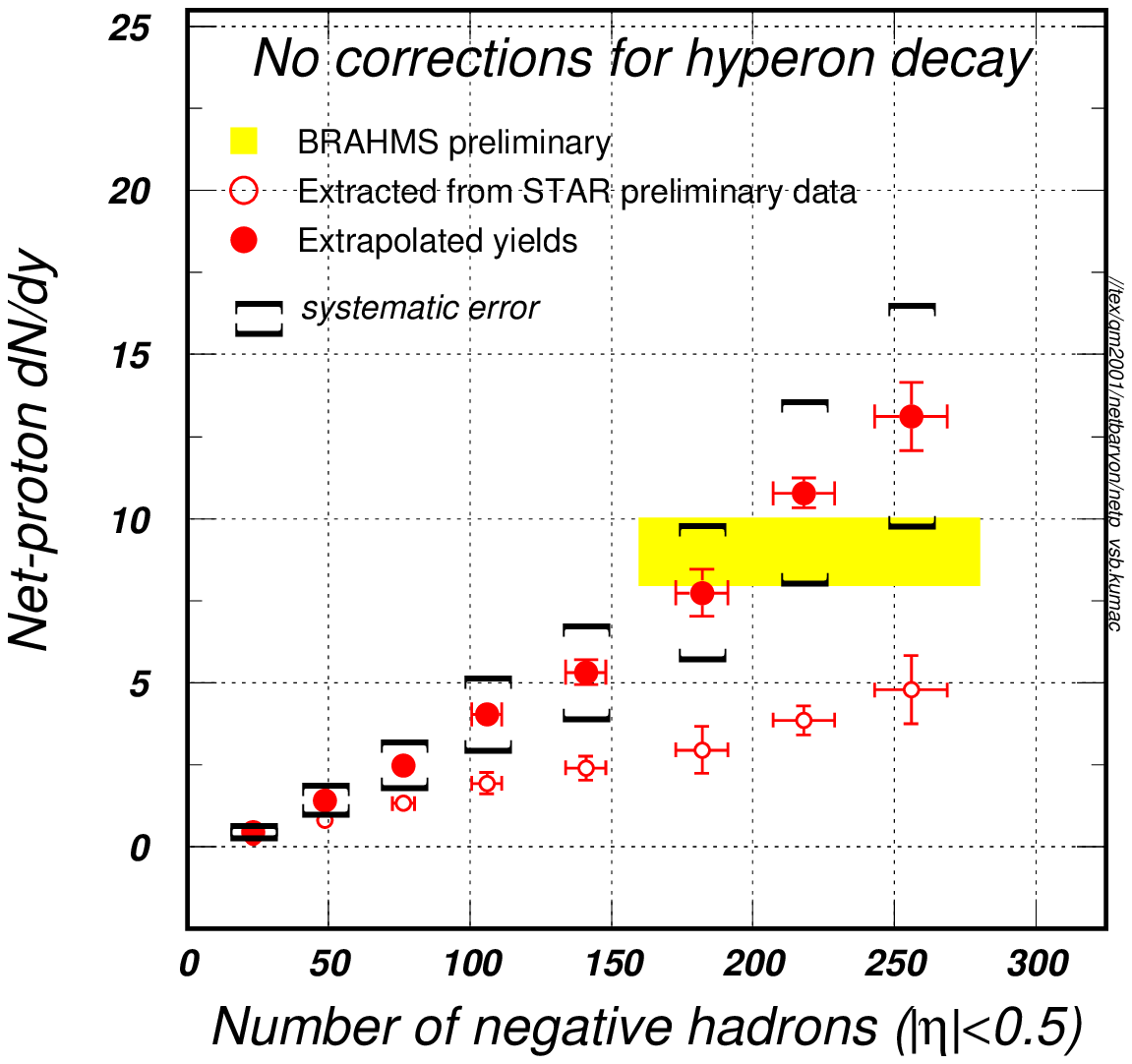,height=8.0cm}}
\end{minipage}

\vspace{-.65cm} \hspace{-0.145cm}  
\begin{minipage}{7.5cm}
  {\small{{\bf Figure 5 }{\it Mid-rapidity $\bar{p}/p$ ratios measured
        in central heavy-ion collisions (filled symbols) and $p+p$
        collisions (open symbols).  The left end of the abscissa is
        the $p$-$\bar{p}$ pair production threshold in $p+p$
        (\rts=3.75~GeV).  }}}
\end{minipage}

\vspace{-2.65cm} \hspace{8.15cm}  
\begin{minipage}{7.25cm}
  {\small{{\bf Figure 6 }{\it Net-protons yields as a function of
        charge particle multiplicity. Open and filled symbols
        represent the yields within 0.3 $\le p_t \le$ 1.0 GeV/c
        and the values extracted from the full transverse momentum
        range, respectively. No hyperon decay corrections have been
        applied to the yields. }}}
\end{minipage}

\vspace{.5cm}

One of the unsolved problems in high energy nuclear collisions is the
baryon transfer that occurs at the early stage of the
collision~\cite{buzsa}. The later dynamic evolution of the system is
largely determined at this moment since the available energy is fixed
at this stage of the collision. On the theoretical side, a novel
mechanism, the non-perturbative gluon junction, was proposed to
address this problem \cite{dima96,vance98}. From the STAR
measured $\bar{p}/p$ ratios \cite{starpbarp} and the STAR preliminary
results of anti-proton yields \cite{harris01}, we have extracted the
net-proton yields at mid-rapidity. The values are shown in Fig.\,6 as
a function of collision centrality.  Open symbols represent the
measured yields within $|y|$ $\le$ 0.1 and 0.3 $\le$ $p_t$ $\le$ 1.0
GeV/c.  The fill symbols show the values extracted from the full
transverse momentum range.  Estimated systematic uncertainties are
shown as caps.  No hyperon decay corrections have been applied as that
requires the knowledge of the yields of hyperons. The hatched bar
represents the BRAHMS preliminary results \cite{vedbeak01}.

One can see that the number of net protons increase as the number of
the negatively charged particles increases. This implies that more and
more protons (baryons) are transfered from the incoming nuclei to
mid-rapidity as the overlap of the two nuclei increases. More data are
needed to study the role of the junction
mechanism~\cite{dima96,vance98} in heavy ion collisions. In
this respect, the rapidity distributions of the net-protons as a
function of collision centrality will be of crucial importance.


\section{Chemical freeze-out: particle ratios}

Recently, much theoretical effort has been devoted to the analysis of
particle production within the framework of statistical models
\cite{pbm9596,becattini98}. These approaches are applied to the
results of both elementary collisions ($e+e^-$,
$p+p$) and heavy ion collisions ($Au+Au$ and $Pb+Pb$) \cite{pbm9596}.
Many features of the data imply that a large degree of chemical
equilibration may be reached both at AGS and SPS energies. The three
most important results are: (i) at high energy collisions the chemical
freeze-out (inelastic collisions cease) occurs at about 160-180 MeV
and it is `universal' to both elementary and heavy ion collisions;
(ii) the kinetic freeze-out (elastic scatterings cease) occurs at a
lower temperature $\sim 120 - 140$ MeV; (iii) the compilation of
freeze-out parameters \cite{cleymans98} in heavy ion collisions in the
energy range from 1 - 200 A$\cdot$GeV shows that a constant energy per
particle $\langle E \rangle / \langle N \rangle \sim 1$ GeV can
reproduce the behavior in the temperature-potential ($T_{ch} - \mu_{_B}$)
plane \cite{cleymans98}.

\vspace{0.1cm} 
\hspace{-.50cm}  
\begin{minipage}{7.cm}
\centerline{\psfig{figure=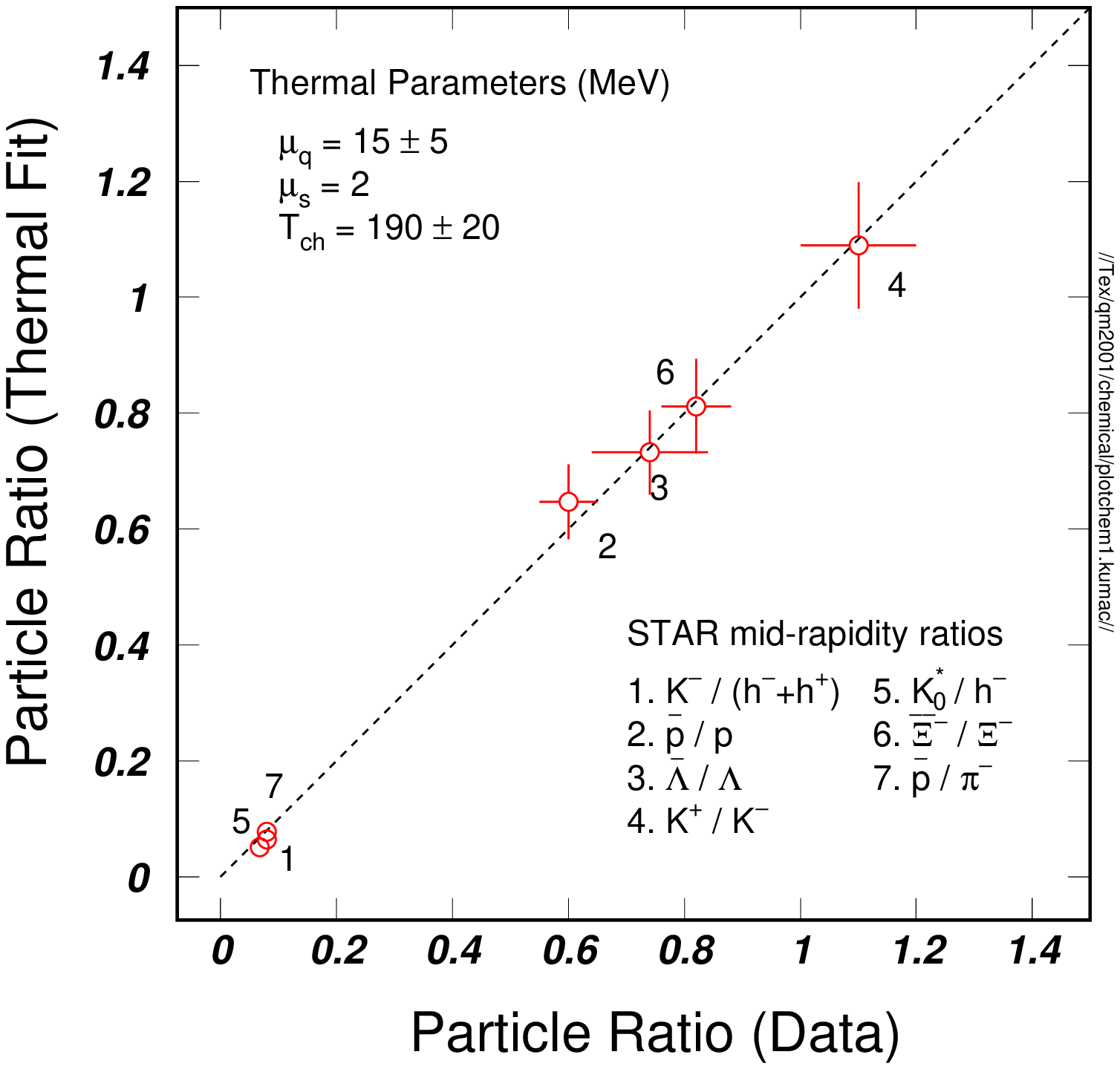,height=7.0cm}}
\end{minipage}

\vspace{-7.65cm}
\hspace{6.5cm} 
\begin{minipage}{8.5cm}
\centerline{\psfig{figure=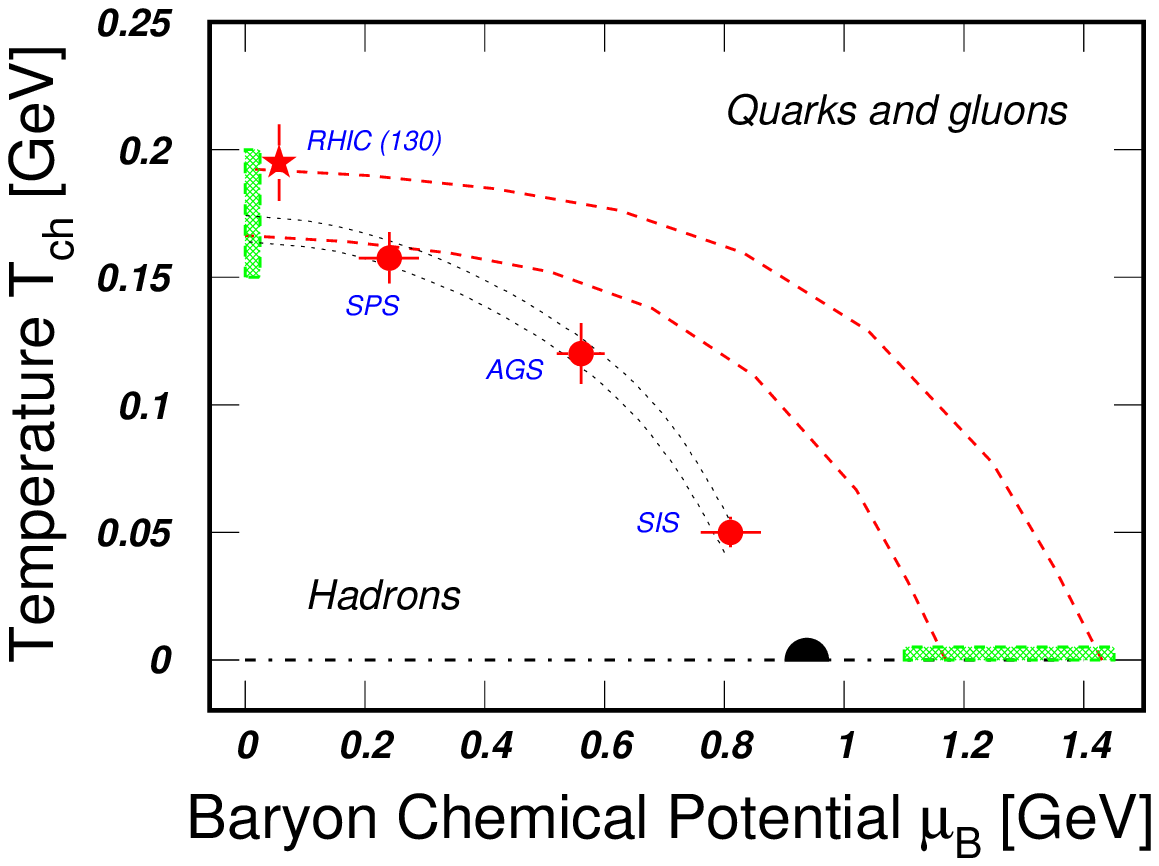,height=8.50cm}}
\end{minipage}

\vspace{-1.cm} \hspace{-0.5cm} 
\begin{minipage}{6.50cm} 
  {\small{{\bf Figure 7 }{\it STAR preliminary particle ratios vs.
        results of thermal model fits. The thermal parameters for
        temperature and chemical potential are $T_{ch} = 190 \pm 20$ MeV
        and $\mu_{_B} = 45 \pm 15$ MeV, respectively. }}}
\end{minipage}

\vspace{-2.85cm} 
\hspace{6.50cm} 
\begin{minipage}{8.30cm} 
  {\small{{\bf Figure 8 }{\it Phase plot $T_{ch}$ vs. $\mu_{_B}$.
        Dashed-line represents the boundary between interactions
        involving hadronic and partonic degrees of freedom.
        Dotted-lines represent the results of ~\cite{cleymans98}.  The
        ground state of nuclei is shown as half circle. }}}
\end{minipage}

\vspace{0.75cm}

The measured mid-rapidity particle ratios~\cite{harris01,canes01} were
fitted with a statistical model~\cite{letessier94} and the results are
shown in Fig.\,7.  The resulting thermal parameters for temperature and
baryon potential are $T_{ch} = 190 \pm 20$ MeV and $\mu_{_B} = 45 \pm 15$
MeV, respectively.  It is not surprising that the temperature
parameter is close to the prediction~\cite{pbm01} for RHIC.

The systematics for chemical parameters is shown in the phase plot in
Fig.\,8 where the dashed lines represent the boundary between
interactions involving hadronic and partonic degrees of freedom.  The
dotted lines represent the results of ~\cite{cleymans98}. Results from
heavy ion collisions at SIS, AGS, and SPS energies are shown as
circles.  The result from collisions at the RHIC energy is shown as
star. The lattice prediction on $T_c$ ($\approx 160$ MeV) is shown at
$\mu_B = 0$ and the baryon density for neutron star is indicated at
the $T_{ch} = 0$.

There are technical caveats in above approaches. The application of a
statistical model requires that the measurement is done over the whole
phase-space, {\it i.e.}, with 4$\pi$ particle yields.  This is because
conservation laws that apply to the collisions are only valid for
global measurements, {\bf{\it not locally}}.  In all experiments the
coverage of the phase-space is limited. Therefore, extracting the
yields measured within limited phase-space to 4$\pi$ yields is almost
impossible, especially for the collider experiments. In addition, we
should point out that the non-local effect is common to all global
conservation laws (baryon number, charge, and so on), not only to the
strangeness.

\section{Summary}

In summary, the most interesting results from collisions at RHIC are
that the system is indeed approaching net-baryon free and the
transverse expansion is much stronger than that from collisions at
AGS/SPS energies.

In order to understand the trend of the collective velocity, an energy
scan between \rts = 20 - 200 GeV, is important. In addition,
systematic studies on the anisotropy parameter $v_2$ and
the transverse momentum distributions of $\phi, \Omega,$ and $J / \psi$ are
necessary as they will help in determining whether the collectivity is
developed at the partonic stage.

 -----
 
 We are grateful for many enlightening discussions with Drs. P.
 Braun-Munzinger, W. Busza, M. Gyulassy, U. Heinz, B. Holzman, P.
 Huovinen, B. Jacak, F. Navarra, S. Panitkin, K.  Redlich, H.G.
 Ritter, K.  Schweda, E.V. Shuryak, R. Snellings, D.  Teaney, F.Q.
 Wang, and X.N.  Wang. This work has been supported by the U.S.
 Department of Energy under Contract No. DE-AC03-76SF00098.




\begin{thebibliography}{9}

\bibitem{vedbeak01} F. Videbaek {\it et al.}, (BRAHMS Collaboration), these proceedings.

\bibitem{zajc01} B. Zajc {\it et al.}, (PHENIX Collaboration), these proceedings.

\bibitem{roland01} G. Roland {\it et al.}, (PHOBOS Collaboration), these proceedings.

\bibitem{harris01} J. Harris {\it et al.}, (STAR Collaboration), these proceedings.
  
\bibitem{friese01} V. Friese {\it et al.}, (NA49 Collaboration), these proceedings.
  
\bibitem{bordalo01} P. Bordalo {\it et al.}, (NA50 Collaboration),
  these proceedings.
  
\bibitem{na50plb} M.C. Abreu {\it et al.}, (NA50 Collaboration), Phys.
  Lett. {\bf B499}, 85(2001).

\bibitem{harry01} H. Appelsh\"aeuser {\it et al.}, (NA45 Collaboration), these proceedings.

\bibitem{blume01} C. Blume {\it et al.}, (NA49 Collaboration), these proceedings.

\bibitem{dress01} A. Drees, these proceedings.
  
\bibitem{steiberg01} P. Steinberg, these proceedings.

\bibitem{panitkin01} S. Panitkin, these proceedings.

\bibitem{julia01} J. Velkovska {\it et al.}, (PHENIX Collaboration), these
  proceedings.

\bibitem{mauel01} Mauel Calder\'on de la S\'anchez {\it et al.}, (STAR
  Collaboration), these proceedings.
  
\bibitem{heinz87} E. Schnedermann, J. Sollfrank, and
  U. Heinz, Phys. Rev. {\bf C48}, 2462(1993).
  
\bibitem{na44_coll_exp} I. G. Bearden {\it et al.}, (NA44
  Collaboration), Phys. Rev. Lett {\bf 78}, 2080(1997); M. Kaneta,
  Ph.D Thesis, (NA44 Collaboration), March (1999).
  
\bibitem{tsukuba} E. Andersen {\it et al.}, (WA97 collaboration), Phys.
  Lett. {\bf B433}, 209(1998).
  
\bibitem{nxuqm96} N. Xu {\it et al.}, (NA44 Collaboration), Nucl.
  Phys. {\bf A610}, 175c(1996).
  
\bibitem{johnson} S. Johnson, B. Jacak, and A. Drees, E. J. Phys. {\bf
    C}, in print, (2001).
  
\bibitem{soffsqm00} S. Soff {\it et al.}, J. Phys. G: Nucl. Part. {\bf
    27}, 449(2001).

\bibitem{pass01} P. Huovinen, these proceedings.

\bibitem{deark01} D. Teaney, these proceedings.
  
\bibitem{passplb01} P. Huovinen, P. Kolb, U. Heinz, P.V. Ruuskanen,
  and S. Voloshin, Phys. Lett. {\bf B503}, 58(2001).
  
\bibitem{starflow} K.H. Ackermann {\it et al.}, (STAR Collaboration),
  Phys. Rev. Lett.  {\bf 86}, 402(2001).
  
\bibitem{snelling01} R. Snellings {\it et al.}, (STAR Collaboration),
  these proceedings.

\bibitem{hsx98} H. van Hecke, H. Sorge, and N. Xu,
Phys. Rev. Lett. {\bf 81}, 5764(1998).

\bibitem{schwinger} J. Schwinger, Phys. Rev. {\bf 82}, 664(1951).

\bibitem{soff} S. Soff, S. Bass, M. Bleicher, L. Bravina, M.
  Gorenstein, E. Zabrodin, H. St\"ocker, and W. Greiner, Phys. Lett.
  {\bf B471}, 89(1999).

\bibitem{bleicher001} M. Bleicher, W. Greiner, H. St\"ocker, and N. Xu,
  Phys. Rev. {\bf C62}, R1901(2000).
  
\bibitem{thews01} R. Thews, M. Schroedter and J. Rafelski, J. Phys.
  {\bf G27}, 715(2001); R. Thews, M. Schroedter and J. Rafelski, LANL
  Preprint hep-ph/0007323 v3, Feb. (2001).


\bibitem{mclerran01} L. McLerran and J. Schaffner-Bielich, LANL
  Preprint hep-ph/0101133, Jan. (2001).

\bibitem{pomeranchuk} I.Ya. Pomeranchuk, Dokl. Akad. Sci. U.S.S.R.
{\bf 78}, 884(1951).

\bibitem{heg} R. Hagedorn, Suppl. Nuovo Cim. {\bf III.2}, 147(1965).

\bibitem{stocker} H. St\"ocker, A.A. Ogloblin, and W. Greiner, Z.
Phys. {\bf A303}, 259(1981).

\bibitem{cramer01} J. Cramer {\it et al.},
 (STAR Collaboration), QM01, Poster Session.

\bibitem{lisa01} M. Lisa {\it et al.},
 (E895 Collaboration), these proceedings.

\bibitem{laue01} F. Laue {\it et al.},
 (STAR Collaboration), these proceedings.

\bibitem{vH83} L. van Hove, Z. Phys. {\bf C21}, 93(1983).
  
\bibitem{navarra92} F. Navarra, M. Nemes, U. Ornik, and S. Paiva,
  Phys. Rev.  {\bf C45}, R2552(1992); Y. Hama and F. Navarra, Z. Phys.
  {\bf C53}. 501(1992).

\bibitem{starpbarp} C. Adler {\it et al.}, (STAR Collaboration),
  Phys. Rev. Lett., in print (2001).
  
\bibitem{beadern01} I. Bearden {\it et al.}, (BRAHMS Collaboration),
  these proceedings.
  
\bibitem{georg01} N. George {\it et al.}, (PHOBOS Collaboration),
  these proceedings.

\bibitem{hirok01} H. Ohnishi {\it et al.},
 (PHEMIX Collaboration), these proceedings.
 
\bibitem{huan01} H. Huang {\it et al.}, (STAR Collaboration), these
  proceedings.
  
\bibitem{buzsa} W. Busza and R. Ledoux, Ann. Rev. Nucl. Part. Sci. {\bf
    38}, 119(1988).

\bibitem{dima96} D. Kharzeev, Phys. Lett. {\bf B378}, 238(1996).
  
\bibitem{vance98} S. Vance, M. Gyulassy, and X.N. Wang, Phys. Lett.
  {\bf B443}, 238(1998).

\bibitem{pbm9596} P. Braun-Munzinger, J. Stachel, J. Wessels, and N.
  Xu, Phys. Lett. {\bf B344}, 43(1995); P. Braun-Munzinger, I. Heppe,
  and J. Stachel, Phys.  Lett. {\bf B465}, 15(1999).
  
\bibitem{becattini98} F. Becattini, M. Gazdzicki, and J. Sollfrank,
  Eur. Phys. J. {\bf C5}, 143(1998); F. Becattini, Z. Phys. {\bf C69}, 485(1996); F.
  Becattini and U. Heinz, Z. Phys. {\bf C76}, 269(1997).

\bibitem{cleymans98} J. Cleymans and K. Redlich, Phys. Rev. Lett. {\bf
    81}, 5284(1998).

\bibitem{canes01} H. Caines {\it et al.}, (STAR Collaboration), these proceedings.

\bibitem{letessier94} J. Letessier, J. Rafelski, and A. Tounsi, Phys.
  Lett. {\bf B328}, 499(1994); M. Kaneta {\it et al.}, (NA44 Collaboration), J.
  Phys. G: Nucl. Part. Phys. {\bf 23}, 1865(1997).
  
\bibitem{pbm01} J. Stachel, Nucl. Phys. {\bf A661}, 226c(1999); P.
  Braun-Munzinger and K. Redlich, private communications, (2001).

\end{thebibliography}
\end{document}